\documentclass[12pt]{article}
\usepackage{tikz}
\usetikzlibrary{matrix,arrows,decorations.pathmorphing,decorations.pathreplacing}
\usepackage{jheppub}
\usepackage{amsmath,amssymb,euscript,array,mathrsfs,appendix,ctable,marvosym}
\usepackage{arydshln}
\usepackage{todonotes}
\usepackage{graphicx}

\newcommand\blank[1]{#1}
\renewcommand\blank[1]{}
\def\Buildrel#1\over#2\under#3{\mathrel{\mathop{\kern0pt
#2}\limits^{#1}_{#3}}}

\newcolumntype{C}[1]{>{\centering\arraybackslash$}m{#1}<{$}}
\newlength{\mycolwd}                                         
\def\s{\sigma}

\def\B0{{\boldsymbol 0}}

\def\Dbarslash{\,\,{\raise.15ex\hbox{/}\mkern-12mu {\bar D}}}
\def\Dslash{\,\,{\raise.15ex\hbox{/}\mkern-12mu D}}
\def\delslash{\,\,{\raise.15ex\hbox{/}\mkern-9mu \partial}}
\def\delbarslash{\,\,{\raise.15ex\hbox{/}\mkern-9mu {\bar\partial}}}

\newcommand{\MAT}[1]{\begin{pmatrix} #1\end{pmatrix}}

\newcommand{\EQ}[1]{\begin{equation}\begin{split} #1
\end{split}\end{equation}}

\def\gg{{\mathfrak g}}
\def\ss{{\mathfrak s}}
\def\s{{\bf s}}
\def\Pneg{P_{<0[\s]}}
\def\Ppos{P_{\geq0[\s]}}

\title{Generalized Integrability and Two Dimensional Gravitation\footnote{
Based on talks given at the ``Conference on Modern Problems in Quantum Field
Theory, Strings and Quantum Gravity'', Kiev, June 8-17 (1992)}}

\author[a]{Timothy J. Hollowood}
\author[a]{Luis J. Miramontes}
\author[b]{Joaqu\'\i n S\'anchez Guill\'en}

\affiliation[a]{Theory Division, CERN, CH-1211 Geneva 23, Switzerland}
\affiliation[b]{Department of Particle
Physics, University of Santiago, E-15706 Santiago de Compostela, Spain}

\abstract{We review the construction of generalized integrable hierarchies of
partial differential equations, associated to affine Kac-Moody algebras,
that include those considered by Drinfel'd and Sokolov. These hierarchies
can be used to construct new models of 2D quantum or topological
gravity, as well as new $\cal W$-algebras.}

\notoc
\begin{document}

\maketitle

\newpage
\section{Introduction}
The combined technology of matrix-models and the double scaling limit
have provided a new way to study two-dimensional quantum gravity theories
\cite{RS},\cite{MM}. This technology has uncovered
an integrable structure underlying such
theories which seems to be of deep significance \cite{DOUG}. The
integrability
is manifested as a classical integrable hierarchy of equations of the
KdV-type, with the partition function of the model being the tau-function of
the hierarchy, subject to an additional constraint known as the string
equation. The flows of the hierarchy are identified with the couplings
to the local operators of the theory which define the background.
For the original one-matrix model the hierarchy is precisely the
original KdV hierarchy, whereas for the $n$-matrix model the hierarchy is
the Drinfel'd-Sokolov $A_n$ generalized KdV hierarchy~\cite{DS}.
The continuum
theories have been identified as pure gravity coupled to various kinds
of matter systems \cite{MAT},\cite{DOUG},\cite{KUT}.

{}From a slightly different perspective, it has been discovered that in the
moduli space of such a model --- spanned by the flows of the hierarchy ---
there exists a very special point for which the theory is topological; in
fact topological gravity coupled to a topological matter theory. These latter
matter theories can be constructed by ``twisting'' $N=2$
superconformal field theories (SCFTs). Here's the rub:
there are many
more known $N=2$ superconformal field theories than known matrix models
\cite{KS}: the $n$ matrix model corresponds to the
 $A_{n-1}$ minimal $N=2$ SCFT. This begs the
question as to whether each $N=2$ SCFT, when twisted
and coupled to topological gravity, will be
related to some integrable hierarchy.
Some progress has been made for the Drinfel'd-Sokolov hierarchies associated
to the $D$ and $E$ algebras \cite{KUT},
\cite{DE} --- although a matrix model has not
been identified in these cases --- which are related to the $D$ and $E$
series of minimal $N=2$ SCFTs. However, even for the exceptional algebras
the string equations appropriate to the hierarchies do not seem to have been
found.

In attempting to find new models of two-dimensional gravity one could try to
generalize the matrix model construction and the double scaling
limit. An alternative approach --- and the one adopted here --- is to
concentrate
on the integrable structure itself, by finding natural generalizations of the
hierarchies and string equations. One could then determine whether these new
models had ``topological points'' and whether these corresponded to some
$N=2$ SCFT. In retrospect, one could then attempt to
discover whether there was an underlying matrix model. This work goes some
way towards this goal by finding a natural candidate for the string
equation; in particular, we find the string equations for all the
Drinfel'd-Sokolov hierarchies including the exceptional algebras, as well as
a more general class of hierarchies.
We then go on to show that these string equations naturally lead to
Virasoro constraints, and more generally ${\cal W}$-algebra constraints,
 on the tau-functions of the hierarchies; again
generalizing the situation for the $A$ and $D$ Drinfel'd-Sokolov
hierarchies. We shall also make some tentative comments regarding the
existence of topological points in the coupling constant space of the more
general models. In this regard, we should
also mention that other kinds of matrix models have been constructed which
seem to describe the topological points of the matrix models \cite{KONT}.
These models seem to have a deep significance for the topological theories and
it would clearly be interesting to know whether they could be generalized
to encompass the more general picture that we discuss in this paper.

The organization  of this paper is the following.
In Section 2 we review the construction of a general class of integrable
hierarchies, which include the Drinfel'd-Sokolov hierarchies as particular
cases. The construction is couched in the language of Kac-Moody algebras,
rather than the pseudo-differential operator
approach. As an aside, we explain how the hierarchies are Hamiltonian, and
hence
how they lead to generalized classical ${\cal W}$-algebras. We go on to
construct a
tau-function formalism of the hierarchies and show they are certain
Kac-Wakimoto hierarchies \cite{KW}. In Section 3 we show
that each of the hierarchies admits a natural ``string equation'' which
leads to Virasoro constraints on the tau-function, generalizing the
situation for the $A_n$ KdV hierarchies. We also point out that there
exist more complicated constraints which, following the conventional wisdom,
should generate ${\cal W}$-algebra constraints. Although a direct proof of
this latter point is still lacking we are able to show that such
${\cal W}$-algebra
constraints are compatible with the integrable hierarchy. In
Section 5 we end with some comments.
\section{Review of Generalized Hierarchies}
In this section we review the construction of a general class of integrable
hierarchies \cite{GEN1},\cite{GEN2},\cite{GEN3},\cite{KW}.
This class includes the
Drinfel'd-Sokolov hierarchies as special cases. Our approach makes use of
the ``matrix'' Lax formalism, as opposed to the scalar form. This is not
because we believe that it is not possible to formulate the hierarchies
using the scalar Lax form, but rather the ``matrix'' formulation introduces
the powerful technology of Kac-Moody algebras. We will go on to show in
following sections that all these hierarchies naturally lead to the
integrable structures like those found in the multi-matrix model, although,
as we have
stressed in the introduction, a matrix-model formulation of these generalized
cases is not yet available. In order that the presentation be reasonably
short we shall refer the reader to refs. \cite{GEN1},\cite{GEN2},\cite{GEN3}
for clarification of any technical points.

We should also mention that there are related works which construct
generalized integrable hierarchies \cite{OTH}.

\section{Zero-curvature Hierarchies}

In refs. \cite{GEN1},\cite{GEN2}, the zero-curvature or Lax formalism was
used to associate a generalized integrable hierarchy
to each affine Kac-Moody algebra $\gg$,
a particular Heisenberg subalgebra $\ss\subset \gg$ (with associated
gradation $\s'$), and an additional gradation $\s$, such that
$\s\preceq\s'$, with respect to a partial ordering. A gradation $\s$ is a
set of non-negative integers associated to the Dynkin diagram of $\gg$
\cite{KAC}, and
we denote the decomposition of $\gg$ with respect to $\s$ as
\EQ{
\gg=\bigoplus_{j\in{\Bbb Z}}\gg_j(\s).
}
The partial ordering is defined by saying $\s\preceq\s'$ if $s'_i\neq0$
whenever $s_i\neq0$. For the definition of a Heisenberg subalgebra we refer
the reader to \cite{KACPET}, however, for our purposes we note that such an
algebra is abelian up to the centre.
In \cite{GEN1}, the construction was undertaken in the loop algebra
(Kac-Moody algebra with zero center) but it was shown in \cite{GEN3}
that the resulting hierarchies
are actually independent of the center --- indeed they are completely
representation independent. In general, there is a flow of
the hierarchy for each element of $\ss$ of non-negative $\s'$-grade,
this is the set $\{ b_j\,,\,\, j\in E\geq0\}$, where the subscripts label the
$\s'$ grade and $E$ is a subset of integers. We note that there exists a
set of rank$(\gg)$ numbers $I$,
$\leq0$ and $<N$, for some integer $N$, where $E=I+N{\Bbb Z}$. At this point,
we also mention that it is possible for there to be a ``degeneracy'' in the
labelling of $E$, that is more than one element for a given grade. In the
following we shall not label this degeneracy explicitly.
The dynamical variables are the components of
the {\it potentials} $q(j)$ in the Lax operators:
\EQ{
{\cal L}_j = {\partial \over \partial t_j} - b_j -q(j) \qquad
j\in E\geq0\,,
\label{Lax}
}
where $q(j)$'s are functions of all the {\it times\/}, $t_m$, on the
intersection
\EQ{
Q(j) \equiv \gg_{\geq0}(\s) \bigcap \gg_{<j}(\s')\,.
\label{spac}
}
In the above, notation like $\gg_{<j}(\s')$ means the subspace of $\gg$
generated by elements with $\s'$-grade less than $j$. In order to ensure
that the flows $t_j$ are uniquely associated to elements of the
set $\{b_j,\,j\in E\geq0\}$ we will also demand that $q(j)$ has no
constant terms proportional to $b_i$ with $i<j$. The integrable hierarchy
of equations is defined by the zero-curvature conditions
\EQ{
\left[ {\cal L}_i,{\cal L}_j \right] = 0\,.
\label{hier}
}

At the moment the hierarchy is to be thought of a set of partial
differential equations in the variables $q(j)$. However,
a hierarchies is usually presented as a set of partial differential
equations in a finite set of variables; we now consider how this is
achieved, and we shall see that there are many --- in general an infinite
number --- of ways to do this. For each {\it regular} element $b_k \in \ss\,
(k>0)$, so that $\gg$ admits the decomposition
\EQ{
\gg = \ss \oplus {\rm Im}\, ( {\rm ad}\, b_k)\,,
\label{decom}
}
we may regard \eqref{hier} as an integrable hierarchy of partial differential
equations on $q(k)$, modulo a gauge symmetry we discuss
below. (In the language of \cite{GEN1} these are the ``type-I'' hierarchies.)
In this case, $q(k)$ are the dynamical variables of the hierarchy,
and one can express all the other functions $q(j)$ for $j\neq k$
in terms of $q(k)$ and its $t_k$-derivatives. In the language of the
matrix models, $t_k$ is a possible choice for $x$, the generalized
``cosmological constant'', and the potential $q(k)$ is the analog of
the functions parameterizing the double scaling limit; for example, $u$
and $\nu$ in the KdV and mKdV cases, respectively. Sometimes, for such a choice
of $b_k$, we shall refer to
${\cal L}_k$ as {\it the\/} Lax operator
$L\equiv{\cal L}_k=\partial/\partial t_k\,- b_k -q(k)$.

The hierarchy exhibits a {\it gauge\/} invariance of the form
\EQ{
{\cal L}_j \rightarrow U{\cal L}_j U^{-1}\,,
\label{gaug}
}
preserving $q(j)\in Q(j)$, where $U$ is a function on the group generated
by the finite dimensional subalgebra given by the intersection
\EQ{
P\equiv \gg_0(\s)\bigcap \gg_{<0}(\s')\,.
\label{gaugespace}
}
Consequently, the equations of the hierarchy are to be though of
as equations on the equivalent classes of $Q(j)$ under the gauge
transformations, which will be parameterized by certain gauge invariant
functions $\tilde q(k)$. The equations of the hierarchy are then of the
form
\EQ{
{\partial \tilde q(k)\over \partial t_j} = F_j \left( \tilde q(k),
{\partial \tilde q(k)\over \partial t_k},{\partial^2\tilde
q(k)\over\partial t_k^2},\ldots \right) \qquad
j\in E\geq0\,.
\label{eqfin}
}
Notice that if $\s\simeq \s'$ then $P=\emptyset$.
We shall refer to
this case as a (generalized) ``modified'' KdV hierarchy, because it
generalizes the standard $sl(2)$-mKdV case. In fact, the auxiliary gradation
$\s$ sets the ``degree of modification'' of the hierarchy: the larger
$\s$ becomes, the more ``modified'' the hierarchy becomes. We shall refer
to the case where $\s$ is the homogeneous
gradation as a (generalized) KdV hierarchy, while other
intermediate cases with
$\s_{\rm hom}\prec \s \prec \s'$ will be called ``partially modified'' KdV
hierarchies (pmKdV). When $\s_2\prec\s_1$ there exists a generalized
Miura transformation taking solutions of the hierarchy with $\s_1$ to
solutions of the hierarchy $\s_2$ \cite{GEN3}.

It is worth making a short comment about the modifications that result from
considering the centre of the Kac-Moody algebra since Drinfel'd and Sokolov
worked in the loop algebra.
The gauge transformations in \eqref{gaug} also include transformations proportional
to the centre of $\gg$. These transformations can be
used to set the possible component of $q(k)$ in the center of $\gg$,
$q_c(k)$, to any arbitrary value, which shows that
it is not a true dynamical degree of
freedom. This, and the fact that $q_c(k)$ cannot contribute to the
evolution of the other components of $q(k)$, is the reason why the
hierarchy is independent of the centre; in fact, the hierarchies are
completely representation independent.

The above construction contains as particular cases the hierarchies
built by
Drinfel'd and Sokolov \cite{DS}, which have been realized in the
analysis of
matrix models \cite{MM},\cite{DOUG} and two-dimensional
Topological gravity \cite{TOP}.
In particular, consider the loop algebra
$\gg = sl(N)^{(1)}$, choosing the defining N-dimensional representation
of $sl(N)$, and the principle Heisenberg subalgebra, whose elements are
$b_j = \Lambda^j$, with  $ j\neq N{\Bbb Z}$, where
\EQ{
\Lambda={\small\MAT{&1&&& \\  &&1&& \\  &&&\ddots& \\  &&&&1 \\  z&&&&} }\,,
}
and zeros elsewhere; $E=\{1,2,\ldots,N-1,\ {\rm mod}\,N\}$ in this
case. If we
choose $x=t_1$, and $\s$ to be the homogeneous gradation, then we obtain
the Drinfel'd--Sokolov KdV hierarchy associated to $sl(N)$, where
$q(1)$ is just a lower triangular matrix, and gauge
transformations are generated by strictly
lower triangular matrices. The ``Drinfel'd-Sokolov gauge'' is
\EQ{
\tilde q(1)= \MAT{0&&&& \\  &0&&& \\  &&\ddots&& \\  &&&0& \\ 
u_1& u_2&\ldots&u_{N-1}&0} \,.
}
Conversely, if $\s$ is chosen to be the principle gradation, then
the resulting hierarchy is the Drinfel'd--Sokolov $sl(n)$ mKdV hierarchy,
where $q(1)$ is just a trace-less diagonal matrix, there being no gauge
freedom in this case.

Along with each integrable hierarchy, there is an associated linear
problem:
\EQ{
{\cal L}_j\Psi=\left[ {\partial \over \partial t_j} - b_j - q(j)\right] \Psi
=0\qquad j\in E\geq0\,,
\label{linear}
}
where $\Psi$ is a function of the $t_j$'s on the Kac-Moody group $G$
formed by {\it exponentiating\/} $\gg$. The zero-curvature conditions
\eqref{hier} are derived as
the integrability conditions of the associated linear problem \eqref{linear}.
If we introduce $\Theta$ via
\EQ{
\Psi=\Theta\exp\left[\sum_{j\in E\geq0}t_jb_j\right],
\label{defgam}
}
then generally $\Theta\in U_-(\s')$, the group formed by exponentiating
$\gg_{<0}(\s')$. In \cite{GEN3}, we have proved that there exists a
gauge-choice, which we call tau-gauge, where $\Theta$ is restricted to be
in $U_-(\s)$, the group generated by $\gg_{<0}(\s)$.
We emphasize that this is {\it not\/} the same as the
Drinfel'd-Sokolov gauge; for example, for the original KdV hierarchy the two
gauges are
\EQ{
\tilde q(1)_{\rm DS}=\MAT{0&0 \\  -u&0} ,\qquad
\tilde q(1)_\tau=\MAT{w&0 \\  w''-(w')^2&-w}\,,
}
where $\prime$ denotes a $t_k$-derivative. From now on we shall assume that
the tau-gauge has been chosen.
An important result from \cite{GEN3} (Theorem 2.4),
which we shall use in the next section is that there is a one-to-one map
from solutions of the associated linear problem, in tau-gauge, of the
form \eqref{defgam}, where $\Theta$ is a function on the subgroup $U_{-}(\s)$, and
solutions of the hierarchy \eqref{eqfin}.

More explicitly one can show
\EQ{
  &\tilde q(j) = P_{\geq0[\s]} \left( \Theta b_j \Theta^{-1} \right)
-b_j \in Q(j) \,, \\ 
& {\partial \Theta\over \partial t_j} = - P_{<0[\s]} \left(\Theta b_j
\Theta^{-1}\right) \Theta \,, 
\label{thetaflow}
}
where $P_{\geq0[\s]}$ is the projector onto non-negative $\s$-grade,
We can use these last two equations to define the hierarchy in a very
symmetrical way with respect to the choice of $t_k$. In this case
all the information is encoded in $\Theta$, and,
once we have chosen $t_k$, we shall understand $\Theta$ to be a function of
$\tilde q(k)$ and its $t_k$ derivatives, the other $\tilde q(j)$
will be given by \eqref{thetaflow}. In the tau-function formalism
(see below) $\Theta$, and hence $\tilde q(k)$, will be understood as
functionals of the tau-functions, and will be the key for connecting
the two formalisms.

\section{Hamiltonian Structures and ${\cal W}$-Algebras}
In this subsection, we briefly describe how the zero-curvature hierarchies
can be cast in a Hamiltonian form. Although this will not be required later,
it is interesting because some of the
Poisson bracket algebras are actually classical ${\cal W}$-algebras, and so our
approach leads to a new classification of such algebras. For related works
on classifying ${\cal W}$-algebras see ref. \cite{WOTH}.

It is possible to set up a Hamiltonian formalism for each distinct choice of
$b_k$ which admits the decomposition \eqref{decom}. The Poisson bracket algebra is
defined on gauge invariant functionals of the variables of the Lax operator
$L$. To define the ``second'' Hamiltonian structure, we first introduce an
inner product on gauge invariant functionals of $q_k$ in the
Kac-Moody algebra:
\EQ{
(\ ,\ )=\int dt_k\langle\ ,\ \rangle,
}
where $\langle\ ,\ \rangle$ is the
Killing form on the Kac-Moody algebra. The functional derivative
$d\varphi$ of a gauge invariant functional $\varphi$ of $q_k$ is defined via
\EQ{
\left.{d\over d\varepsilon}\varphi[q_k+\varepsilon
r]\right\vert_{\varepsilon=0}=\left(d\varphi,r\right),
}
for all functions $r\in Q_k$. The second Hamiltonian structure is then
\EQ{
\left\{\varphi,\psi\right\}=\left(d\varphi_0,[d\psi_0,L]\right)
-\left(d\varphi_{<0},[d\psi_{<0},L]\right),
}
where $L\equiv{\cal L}_k$ and the subscripts indicate projections onto
$\s$-grade. It can be shown that this Poisson bracket algebra is actually a
classical ${\cal W}$-algebra, generalizing the situation for the Drinfel'd-Sokolov
hierarchies. In particular, consider the case when $\gg=sl(3)^{(1)}$. In
this case there are three inequivalent Heisenberg subalgebras (the Weyl group
of $sl(3)$ has three conjugacy classes). Choosing the principle Heisenberg
subalgebra leads to the Drinfel'd-Sokolov $sl(3)$ KdV hierarchy, with the
conventional $W_3$-algebra, whilst choosing the homogeneous Heisenberg
subalgebra leads to a $sl(3)$ generalization of the non-linear Schr\"odinger
hierarchy; in this case the Poisson bracket algebra is the $sl(3)^{(1)}$
Kac-Moody algebra. The third choice leads to a distinct hierarchy and a
Poisson bracket algebra which is the $W_3^{(2)}$-algebra first discussed in
ref. \cite{PB} (for details see \cite{GEN2} and also \cite{WTWO}).

It turns out that hierarchies based on untwisted Kac-Moody algebra, and when
$\s=\s_{\rm hom}$, admit an additional inequivalent Hamiltonian structure:
the ``first'' Hamiltonian structure. For details of this we refer the
reader to ref. \cite{GEN2}.

\section{The Kac-Wakimoto Hierarchies}

In the construction of Kac and Wakimoto \cite{KW}, the dynamical variables
are the ``tau-functions'', which satisfy a set of bi-linear differential
equations also known as Hirota equations \cite{HIR}. As we have shown
in ref. \cite{GEN3}, and briefly explain
here, in certain cases it is possible
to relate the hierarchies defined in the Lax formalism with those
defined in the tau-function formalism. In these cases, we can write the
variables $\tilde q(j)$ in the Lax operator as a certain combination
of tau-functions.
To make the connection explicit, it will be necessary to consider the
associated linear problem \eqref{linear}.

The tau-function $\tau_{\s}$ associated to an integrable
highest weight representation\footnote{We label an integrable representation by
its Dynkin labels $\s$.} $L(\s)$ of an affine Kac-Moody algebra
$\gg$ is characterized by saying that it lies in the $G$-orbit of the
highest weight vector $v_{\s}$; $G$ being the group associated to $\gg$.

Let $\{u_i\}$ and $\{u^i\}$ be dual basis of the larger algebra
$\gg$ with respect to the non-degenerate bi-linear
inner product $(\cdot|\cdot)$.\footnote{In fact to have a non-degenerate
bi-linear form one has to consider the larger algebra formed by appending
the derivation to $\gg$.} It can be
shown \cite{KW},\cite{PK} that $\tau_{\s}$ lies in the G-orbit of $v_s$ if and
only if
\EQ{
\sum u_j \otimes u^{j} \left( \tau_{\s} \otimes \tau_{\s} \right) =
( \Lambda_{\s} | \Lambda_{\s} ) \tau_{\s} \otimes \tau_{\s}\,,
\label{hirota}
}
where $\Lambda_{\s}$ is the eigenvalue of $\gg_0$ on $v_{\s}$.

Given two tau-functions $\tau_{\s_1}$ and
$\tau_{\s_2}$ then $\tau_{\s}\equiv \tau_{\s_1}\otimes \tau_{\s_2}$ is
a new tau-function with $\s = \s_1 + \s_2$, corresponding to the
representation $L(\s)$ realized as the highest component in the tensor
product $L(\s_1)\otimes L(\s_2)$. Therefore
\EQ{
\tau_{\s} = \bigotimes_{i=0}^{r} \left[ \tau_{i}^{\otimes s_i} \right]\,,
\label{fact}
}
where $\s=(s_0,s_1,\ldots,s_r)$ and
$\tau_i$ is the tau-function  corresponding to the fundamental
representation with Dynkin labels $s_j=\delta_{ij}$.
The conditions
\eqref{hirota} lead to an integrable hierarchy of partial differential equations
when the representation
is  of the ``vertex type'', so that it is carried by the Fock space of a
Heisenberg subalgebra of $\gg$. We shall only consider the case when $\gg$ is
an untwisted affinization of a simply-laced finite Lie algebra $g$.
A fundamental representation with $s_i=\delta_{ij}$ can be
realized in this way if the Kac label associated to the $j^{\rm th}$ node
of the Dynkin diagram has $k_j=1$.
In this case, all the elements of $\gg$ are differential
operators on the representation (Fock) space ${\Bbb C} [x_j]\otimes
{\cal V}$: the center is $c=1$, the elements of the Heisenberg subalgebra
$\ss$ are
\EQ{
b_j = \begin{cases}{\partial \over \partial x_j} \,,& j\geq0  \\ 
- {j\over N}x_{-j}\,,& j<0\,,\end{cases}  
\label{modes}
}
the other elements of $\gg$ are the modes of vertex operators, and the
derivation $d_{\s}$ is, up to a constant, given by the Sugawara construction.
The ``zero-mode'' space ${\cal V}$ is the tensor product of a
finite dimensional vector space $V$ and a space ${\Bbb C}(Q)$, where $Q$
is a certain projection of the root lattice of the finite Lie algebra $g$.
When we come to consider the connection with the zero-curvature hierarchies
we shall only require the vertex operator construction in the restricted
case when the auxiliary vector space $V$ is trivial,
dim$(V)=1$, and so we shall not dwell on its construction.

When realized on the Fock space, the equations \eqref{hirota} are
precisely bi-linear Hirota equations for the functions
$\tau_{i}^{(\beta)} (x_j)$, which are the projections onto a basis for
${\Bbb C}(Q)$:
\EQ{
\tau_i(x_j;x_0) = \sum_{\beta\in Q} \tau_{i}^{(\beta)} (x_j) {\rm e}^{\beta
\cdot x_0} \,.
\label{compon}
}
Notice that, even though the realizations of each level-one
representations based on different Heisenberg subalgebras are isomorphic,
they lead to different Hirota equations and hence different
integrable hierarchies. The inequivalent Heisenberg subalgebras of $g^{(1)}$
are in one-to-one correspondence with the conjugacy classes of the Weyl
group of $g$ \cite{KACPET}.

In general, one can construct Hirota equations for the representations
$L(\s)$ where $s_i\neq0$ only if $k_i=1$, by considering tensor
products of the above vertex representations.

\section{The Connection between the Formalisms}

In this subsection we shall explain how, in certain cases, the Kac-Wakimoto
hierarchies are precisely the tau-function versions of the zero-curvature
hierarchies.

The connection between both constructions is given by Theorem
5.1 of \cite{GEN3}, generalizing the treatment in ref. \cite{IMB}.
There exists a map from solutions of the Kac-Wakimoto
hierarchies associated to the data $\{g,\ss,\s\}$ (with the gradation
associated to the Heisenberg subalgebra $\ss$, denoted $\s'$,
satisfying $\s\preceq \s'$, and $\s$ having $s_i\neq0$ only if $k_i=1$) and a
zero-curvature hierarchy associated to the data $\{ g^{(1)}, \ss,
\s;b_k\}$, given by
\EQ{
\Theta^{-1} \cdot v_{\s} = { \tau_{\s} \left(x_j + t_j\right) \over
\tau_{\s}^{(0)} (t_j)}\,,
\label{connection}
}
where $\Theta\in U_{-}(\s)$ gives $\tilde q(k)$ via \eqref{thetaflow}, and
$\tau_{\s}^{(0)}$ is the component of the tau-function with zero
``momentum'' in \eqref{compon}. We emphasize that the vertex operator
representation in
\eqref{connection}\ is defined by the Fock-space of the $x_j$'s: the $t_j$'s
are to be thought of as auxiliary parameters. We also note that the
condition $\s\preceq\s'$ ensures that the auxiliary vector space $V$ is
trivial. We also point out
the important fact that $U_-(\s)$ is faithful on $v_\s$.
\section{The Generalized String equation}
One of the main results of the random surface or matrix model
formulation of two-dimensional gravity is that it is described
by an integrable hierarchy of partial differential equations, of the KdV-type,
supplemented with an additional condition, known as the ``string
equation'' \cite{DOUG}.
The idea is that the string equation picks out a unique solution
of the hierarchy, although, in certain cases it appears that the solutions
actually exhibit ``non-perturbative'' ambiguities which are not fixed by the
integrable structure. A discussion of this point would take us too far away
from our substantive aim, and so we will not pursue it. This integrable
structure is already present, at a
basic level, in the matrix model describing the discrete random
surface system \cite{TODA}, and it is preserved by the continuum limit
--- the so-called ``double scaling limit''. Let us summarize how the
integrable structure arises in the well-known case of a multi-matrix
model with chain-like interactions, which describes the coupling of
minimal conformal matter to two-dimensional gravity. The matrix model
can be reduced to an integral over the eigenvalues of the matrix,
$\lambda$, and the basic idea, due to Douglas \cite{DOUG}, is that it
is possible to make the double scaling limit on $\lambda$ and
$\partial/\partial\lambda$ seen as operators acting on certain space
of functions; in particular, acting on orthogonal polynomials. The
final result is that the insertion of the eigenvalue $\lambda$ is
represented by a differential operator in the cosmological constant,
$Q$, and that of $\partial/\partial\lambda$ by another operator, $P$,
both of them acting on certain function $\Psi$. Therefore, we have the
two equations
\EQ{
L\Psi \equiv \left(\hat\lambda -
Q\right)\Psi=0\,,\qquad \left({\partial\over \partial \hat\lambda} -
P \right)\Psi =0\,,
\label{PQ}
}
where $\hat\lambda$ is the double-scaled eigenvalue. The important
result is that $L$ is precisely the Lax operator of the
relevant hierarchy; in this case, the Drinfel'd-Sokolov KdV hierarchy
corresponding to $sl(n) $, where $n-1$ is the number of matrices.
In the original argument of Douglas, $\Psi$ is a scalar function,
$L$ is a differential operator of order $n$, and the hierarchy is
described in the scalar Lax formalism. Nevertheless, in the
general case, and depending on the details of
the double-scaling-limit of the orthogonal polynomials,
the hierarchy is given more directly in the
``matrix'' Lax formalism \cite{MOORE},\cite{ISO};
in which case $L$ is a first order differential operator
written in terms of the associated loop algebra. In any case,
$\hat\lambda$ plays the r\^ole of a spectral parameter. Finally,
the compatibility condition for the two differential equations \eqref{PQ},
\EQ{
\left[ L, {\partial\over \partial\hat\lambda} - P  \right] =0\,,
\label{seqmm}
}
is just the string equation, which can be understood as the translation
of the identity $[\lambda, \partial/ \partial \lambda] =-1$ into the
language of the integrable hierarchy.

Of course, there are many more matrix models corresponding
to different matter systems coupled to 2D gravity than those which
have been exactly solved using orthogonal polynomial techniques
\cite{MM},\cite{DOUG}. But, correspondingly, there are also many more
integrable systems that generalize the
Drinfel'd-Sokolov hierarchies, and which could also describe interesting
physical systems coupled to gravity in a similar way. Moreover, it is
known that
the matrix models also describe models of topological gravity coupled to
topological conformal field theory, the latter being constructed from
twisted $N=2$ SCFTs. Again there are many
more $N=2$ SCFTs than the ones which are related to the known
multi-matrix models, suggesting that a more general class of hierarchy and
string equation may be required.

At the very least, if a hierarchy is to describe a model of two-dimensional
gravity then it must admit a generalization of the
string equation \eqref{seqmm}.
Our purpose in this section is to show that the generalized class
of integrable hierarchies discussed in the previous section do indeed
admit a ``string equation'', where we understand by that a
condition of the form \eqref{seqmm}, compatible with
all flows of the hierarchy:
\EQ{
\left[{d\over dz} + S ,  L\right]=0\,,\qquad
{\partial\over\partial t_j} \left[{d\over dz} + S , L\right]=0\,,
\label{seqA}
}
for any $j\in E\geq0$. In addition, when these
hierarchies are related to those constructed of Kac and Wakimoto using
the tau-function formalism \cite{KW},\cite{GEN3}, we will show that the
string equation induces a set of linear constraints on the
tau-functions that may be expressed in terms of the generators of a
$\cal W$-algebra.

Let us consider an integrable hierarchy defined by the zero-curvature
conditions \eqref{hier} in terms of $\{\gg,\ss,\s;b_k\}$, where $\gg$ is the
untwisted affinization of a simply-laced Lie algebra $g$.
For the moment, shall choose the loop algebra representation of
$\gg$, ${\Bbb L}(g) = {\Bbb C} (z,z^{-1}) \otimes g$, safe in the
knowledge that the construction of Section 2 was completely
representation independent.

The possibility of imposing additional constraints like \eqref{seqA} to the
solutions of the hierarchy is related to the existence of additional
non-commuting flows which are, nevertheless, compatible with the
hierarchy. We shall find that these flows are naturally related to an
action of a subalgebra of the Virasoro algebra. This generalizes the
well-known existence of Virasoro deformations of a large class of
integrable hierarchies that includes the KP (and its important reductions
to Drinfel'd-Sokolov KdV's), the NLS, and the Toda lattice
hierarchies \cite{DEFOR}. The formalism is also closely related to
the subject of {\it iso-monodromic\/} deformations \cite{ISO}, but our
approach is closer to that of \cite{SEMI}.

To construct the Virasoro action we first define the quantities $S_n$ via
\EQ{
z^{n+1} {d\over dz} + S_n = \Theta\tilde s_n \Theta^{-1}\,,
\label{prtwoo}
}
with
\EQ{
\tilde s_n =
{z^{n}\over N} \vec d_{\s'} - \sum_{j\in E\geq0} {j\over N}
t_j z^n b_j + \omega_n\,,
\label{prtwoone}
}
where $d_{\s'}$ is the derivation
corresponding to the $\s'$-grading, the arrow indicates that
$\vec d/dz$ acts like $\vec d/dz f(z)\equiv (df(z)/dz) + f(z)d/dz$,
and $\omega_n$ is an arbitrary constant element of $\ss$. In the above
$\Theta$ is related to the hierarchy via \eqref{thetaflow}.

The Virasoro action is then given by
\EQ{ l_n L = \left[z^{n+1} {d\over dz} +
P_{\geq0[\s]}\left(S_n\right) , L\right] =
-\left[P_{<0[\s]}\left(S_n\right), L\right]\,,
\label{Virlax}
}
however, it is important that the action is only valid if $n\geq-1$, for a
KdV-type hierarchy (for which $\s=\s_{\rm hom}$), and $n\geq0$ for
 a (p)mKdV hierarchy (for which $\s_{\rm hom}\prec\s$.

The importance of the Virasoro flows, subject to the constraints on $n$,
is that they commute with all
the other flows of the hierarchy:
\EQ{
\left[{\partial \over \partial t_j}, l_n\right] =0\,.
\label{comiB}
}

In contrast, the Virasoro flows --- as the name implies ---
do not commute, rather they generate a Virasoro
algebra: $[l_m,l_n]=(m-n) l_{m+n}$. Notice that the Virasoro
flows are not sensitive to a possible central extension of this Virasoro
algebra because they are defined only for
$n\geq -1$ or $n\geq0$.

The existence of the Virasoro flows allows one to impose additional
conditions of the form \eqref{seqA}, which are consistent with the hierarchy.
In fact, we can check that the condition
\EQ{
\Pneg \left(S_n\right) =0\qquad n\geq-1\quad{\rm or}\quad0,
\label{cond}
}
is compatible with the hierarchy, in the sense that it is preserved by the
flows. Obviously, \eqref{cond} induces a constraint on the Lax operator of the
form \eqref{seqA}, and hence any of them can be chosen to be the ``string equation''.
Nevertheless, the results obtained with
matrix models suggest that one takes the string equation to be the $n=-1$
constraint, for a KdV hierarchy, and $n=0$ otherwise; indeed, these choices
are the most restrictive.

Therefore, we choose the string equation to be
\EQ{ \Pneg\left(S_{-1}\right) =0\,,
\label{seqKdV}
}
if the hierarchy is of the KdV type, and
\EQ{ \Pneg\left(S_{0}\right) =0\,,
\label{seqmKdV}
}
otherwise, {\it i.e.\/} if it is (partially) modified.

We can also write the string equation
in terms of the Lax operator and compare with \eqref{seqA}.
The resulting condition for hierarchies of the KdV type is
\EQ{
l_{-1}L=\left[{d\over dz}+P_{\geq0[\s]}(S_{-1}),L\right]=0\,,
}
or
\EQ{
\sum_{j\geq N} {j \over N}t_j {\partial L \over \partial t_{j-N}} =
{d L\over dz} + \sum_{0\leq j<N}{j\over N}
\left[t_j z^{-1} b_j,L\right]\,.
\label{seqLKdVB}
}
We emphasize that the result is only applicable for tau-gauge; for other
gauges one must add a compensating gauge transformation to the right-hand
side. The above string equation has the form of \eqref{seqA}, indeed it agrees with
the previously known results
\cite{MOORE},\cite{ISO},\cite{DOUG},\cite{KUT}, but it is also valid
for all the Drinfel'd-Sokolov hierarchies, including the exceptional
algebras, as well as our generalizations.
The corresponding condition on the Lax operator for the (partially) modified
hierarchies is
\EQ{
\left[z{d\over dz} - \sum_{j\in E\geq0} {j\over N} t_j \Ppos\left(
\Theta b_j \Theta^{-1}\right) + {1\over N} \delta_{\s'}\cdot H
+ \alpha b_0\,,\, L\right] = 0\,.
\label{seqLmKdV}
}
In this case the string equation does not have the form of \eqref{seqA}, however, one
can verify in the case of $g=A_n$, that the string equation is that
found in the unitary matrix model \cite{UNIT}, or the two-arc sector of the
Hermitian matrix model \cite{TAM} and also in \cite{POOP}.

\section{The String Equation and the Tau-Functions}

The next step in our story is to show that the string equation leads to
Virasoro constraints on the tau-functions, generalizing the situation
for the multi-matrix models and the $sl(n)$ KdV hierarchies of Drinfel'd and
Sokolov. In fact, it has been conjectured --- but as far as we known not
yet proven in general --- that the string equation induces an infinite set of
constraints that generate a classical $\cal W$-algebra, of which the
Virasoro constraints form a subalgebra \cite{WCONST}.
In particular, for the
Drinfel'd-Sokolov KdV hierarchy associated to $sl(n)$, they generate the
classical ${\cal W}_{n}$-algebra
(${\cal W}_2$ is the Virasoro algebra).

In the general case, under
discussion here, the string equation also induces an
infinite set of constraints. The crucial observation is that
$S_{n}= z^j S_{n-j}$ for any integers $n$ and $j$,
therefore, the string equations \eqref{seqKdV} and \eqref{seqmKdV}
induce the following set of constraints:
\EQ{
\Pneg\left[ z^k( S_{-1})^{m}
\right]=0\qquad m\geq1\,,
\label{genconst}
}
where $k\geq0$ if the hierarchy is of the KdV type, and $k\geq m$ if the
hierarchy is (partially) modified. The constraints
\eqref{genconst} with $m=1$ will lead to Virasoro constraints on the tau-function.
In contrast the other constraints with $m\geq2$ are not associated to
additional symmetries of the hierarchies because $(S_{-1})^{m}$ is no longer
in the loop algebra, however, they also lead to constraints on the
tau-function.

The constraints \eqref{genconst}
derived from the string equations \eqref{seqKdV} and \eqref{seqmKdV} are
expected to induce a
set of linear constraints on the tau-functions when the hierarchy
\eqref{hier} is related by \eqref{connection} to one of those constructed in \cite{KW}.
These hierarchies are defined in terms of level one vertex operator
representations of $\gg=g^{(1)}$, while the set of constraints
\eqref{genconst} are written in terms of the loop algebra ${\Bbb L}(g)$ without
central extension. Therefore, the first step is to
consider how the inclusion of the centre modifies the treatment in the last
subsection. The previous discussion can be repeated with the
 following substitutions
\EQ{
z^{n+1} {d\over dz} \rightarrow - L_n -\sigma \delta_{n,0}\qquad
{z^{n} \over N} d_{\s'} \rightarrow - L_{n}^{\s'} - \eta \delta_{n,0}\,,
\label{changes}
}
where $\sigma$ and $\eta$ are two arbitrary constants that have been
included  for
generality. The definition of the additional Virasoro flows is the same
as before:
\EQ{
& S_n - L_n -\sigma \delta_{n,0}
= \Theta \hat s_n \Theta^{-1} \in g^{(1)}  \\ 
& \hat s_n = L_{n}^{\s'} + \sum_{j\in E\geq0} {j\over N} t_j b_{j+nN}
- b_{nN}    \\ 
& \qquad + c\Big(
{1\over 2} \sum_{j,k\in E\geq0 \atop j+k = -nN}
{jk\over N^2} t_j t_k +  \alpha nN t_{N}\delta_{n+1,0}\Big) + \eta
\delta_{n,0} \,, 
\label{newSn}
}
where the contribution proportional to the center has been fixed by
the identity
\EQ{
\left[ \hat s_n, {\partial\over \partial t_j} - b_j\right]=0\,.
\label{newcero}
}
In \eqref{newSn} $\alpha$ is some constant.
Notice that $S_n$ is a well-defined element of
$g^{(1)}$ because $L_{n}-L_{n}^{\s'} \in g^{(1)}$.
It is straightforward to check that
the Virasoro flows, defined by \eqref{newSn}, are consistent
with the hierarchy and satisfy the commutation relations of the
Virasoro algebra.

Moreover, the conditions \eqref{cond} are still consistent with the hierarchy
because of the identity \eqref{newcero}, and correspond to the generalized
constraints \eqref{genconst} with $m=1$. They ensure that the solutions of the
string equation are constant along the Virasoro flows.

Let us concentrate on the level one vertex operator
representations, which admit a formulation in terms of tau-functions.
In that case, the Virasoro generators are defined in terms of the Heisenberg
subalgebra $\ss$ as
\EQ{
L_{n}^{\s'} = {1\over 2} \sum_{j\in E} :b_j b_{nN-j}:+{1\over4}\sum_{j\in
I}{j\over N}\left(1-{j\over N}\right)\delta_{n,0}\,,
\label{Lnn}
}
and are second order differential operators in $\{x_j\}$
acting on the representation Fock space. Let us distinguish
the dependence on $x_0$ by introducing $p = \partial/\partial x_0$,
and write the Virasoro generators as
\EQ{
L_{n}^{\s'} \equiv L_{n}^{\s'}\left({\partial\over \partial x} ,
x,p\right)\,,
}
where $x$ means $\{x_j\}$ with $j\neq0$.
It is straightforward to check that, in this representation, $\hat s_n$
is precisely
\EQ{
\hat s_n = L_{n}^{\s'} \left({\partial\over \partial x} , x+ t,
p - \alpha\right) - \left({\alpha^2\over 2}-
\eta\right)\delta_{n,0}\,.
\label{snln}
}
Notice that this form is a direct consequence of \eqref{newcero}, because
\EQ{
{\partial \over \partial t_j} - b_j \equiv
{\partial \over \partial t_j} - {\partial \over \partial x_j} \qquad
j\in E\geq0\,.
\label{trick}
}
Therefore, the final expression for $S_n$ in the level one vertex
operator representation is
\EQ{
S_n = \Theta \left[ L_{n}^{\s'} \left( {\partial\over \partial x},
x+t , p-\alpha\right)  \right] \Theta^{-1}
-L_n -\left({\alpha^2\over2}-\eta+\sigma\right) \delta_{n,0} \,.
\label{Snfin}
}

Now let us consider \eqref{connection}, which gives the relation between the
dynamical variables of the Lax formalism and the tau functions, and
act with \eqref{cond} on $v_{\s}$, the highest weight vector of $L(\s)$:
\EQ{
  \Pneg (S_n) \cdot v_{\s} & =f(t_j)  S_n\cdot v_{\s}  \\ 
&= {f(t_j)\over \tau_{\s}^{(0)}(t_j)} \Theta\left[
L_{n}^{\s'} \left( {\partial\over
\partial x}, x + t , p - \alpha \right)
- \mu \delta_{n,0}\right] \tau_{\s} (x + t)\,,  
\label{viratau}
}
where $f(t_j)$ is the eigenvalue of $P_{0[\s]}(S_n)$ acting on $v_{\s}$
and
\EQ{
\mu=\sigma-\eta+{\alpha^2\over2}+\delta\,,
}
where $\delta$ is the eigenvalue of $L_0$ on $v_\s$.

Therefore, the condition \eqref{cond} implies
\EQ{
\left( L_{n}^{\s'} - \mu \delta_{n,0}\right)
{\rm e}^{-\alpha\cdot x_0} \tau_{\s} =0 \,,
\label{viratauB}
}
{\it i.e.\/} a Virasoro constraint on the tau-function. Notice that since
$\eta$ and $\sigma$ are arbitrary the eigenvalue of $L_0^{\s'}$, $\mu$ is
an arbitrary constant.

Recall that the representation $L(\s)$ is realized as the highest component
in the tensor product of level-one representations. Hence, the Fock space
is a tensor product
of Fock spaces, the tau functions is also a tensor product
of elementary tau-functions \eqref{fact}, and the Virasoro generators can be
split into a sum of generators acting on each copy of the
$N_\s$ Fock spaces:
\EQ{
L_{n}^{\s'} = \sum_{i=1}^{N_\s} L_{n}^{\s'} \left( {\partial\over \partial
x^{(i)}}, x^{(i)}, p^{(i)} \right)\,.
}
Therefore, the Virasoro constraints \eqref{viratauB} can be written in terms
of the elementary tau-functions $\tau_i=\sum_{\beta\in Q}
\tau_{i}^{(\beta)} (x) {\rm e}^{\beta\cdot x_0}$, \eqref{compon}, corresponding to
the fundamental representations $s_j=\delta_{ij}$.

The final result for the Virasoro
constraints implied by the string equation corresponding to the
hierarchy associated to the data $\{g^{(1)},\ss,\s;b_k\}$
(with $\s'$ the gradation associated to $\ss$) is:
\EQ{
L_{n}^{\s'} \left({\partial\over \partial x}, x,\beta - \alpha_i
\right)\cdot \tau_{i}^{(\beta)} (x) = \mu_i \tau_{i}^{(\beta)}(x)
\delta_{n,0} \qquad n\geq0\,,
\label{VirmKdV}
}
if $s_i\neq 0$ ($k_i=1$) for hierarchies of the (p)mKdV type, and
\EQ{
L_{n}^{\s'} \left({\partial\over \partial x}, x,\beta - \alpha
\right)\cdot \tau_{0}^{(\beta)} (x) = 0 \qquad n\geq-1\,,
\label{VirKdV}
}
for hierarchies of the KdV type, where there is a unique
tau-function~\cite{KW},\cite{GEN3}.
Notice that the value of $\mu$
is fixed by the commutation relation $[L_{1}^{\s'} , L_{-1}^{\s'} ] =
2 L_{0}^{\s'} $
only in the KdV case, in contrast to the (p)mKdV hierarchies where it
is arbitrary and, in principle, different for each elementary
tau-function.

We can write the Virasoro constraints in terms of a, generally twisted,
rank$(g)$-dimensional scalar field. To be more
specific, the scalar field is constructed out the Heisenberg subalgebra:
\EQ{
\phi(z)=-ix_0-i\log\,z{\partial\over\partial x_0}+i
\sum_{j\in E>0}
\left({N\over j}z^{-j/N}{\partial\over\partial x_j}-x_jz^{j/N}\right){\boldsymbol
e}(j\ {\rm mod}\,N)\,,
\label{TSF}
}
where the vectors ${\boldsymbol e}(k)$, for $j\in I$,
span a rank$(g)$-dimensional space with inner product
\EQ{
{\boldsymbol e}(k)\cdot{\boldsymbol e}(N-j)=\delta_{jk}\,,
}
(as before we do not explicitly label any degeneracies in $I$). Actually,
the scalar field is twisted by an element of the Weyl group: this illustrates
the connection between the Heisenberg subalgebras of $g^{(1)}$ and the
Weyl group of $g$.
The Virasoro generators in \eqref{VirmKdV} and \eqref{VirKdV} are then the modes of the
``stress-tensor''
\EQ{
T(z)=-{1\over2}:\partial_z\phi\cdot\partial_z\phi:-i\alpha\cdot
\partial_z^2\phi\,.
}
This makes it clear that the freedom present in the parameter $\alpha$ leads
to a Feign-Fuchs like modification of the ``stress-tensor''.

It is worth clearing up a possible point of confusion. We have already seen
in Section 2.2 that the hierarchies are Hamiltonian and that one of the
Poisson bracket algebras is a classical ${\cal W}$-algebra containing as a
subalgebra the Virasoro algebra. However, this Virasoro is not related ---
at least in any obvious way --- with the Virasoro action discussed in this
section.

\section{${\cal W}$ Constraints}

So far, as a consequence of the string equation, we have obtained a set
of Virasoro constraints on the tau-function of the hierarchy, which
are a consequence of \eqref{genconst} with $m=1$. Let us now concentrate on
the other constraints with $m>1$. As we have pointed out before, they
do not follow from any symmetries of the hierarchy, in the same way as the
Virasoro action, because they do not involve
elements of $g^{(1)}$, but rather its universal enveloping algebra. The
natural guess, taking into account the results obtained with the
hierarchies associated to matrix models \cite{WCONST}, is that they correspond
to additional $\cal W$-algebra constraints. A strong argument in this
direction is that they have to be consistent with the Virasoro constraints,
and so form a closed algebra with the Virasoro algebra as a subalgebra.
Although there have been a number of attempts to prove directly that ${\cal
W}$-algebra constraints arise, as yet there is no proof in the general case.
We shall not be able to remedy the situation here, rather we shall limit
ourselves to some observations

First of all, we shall show that the ${\cal W}$-algebra constraints are
compatible with the hierarchy. The generators of the $\cal W$-algebra
are constructed out of the Heisenberg subalgebra $\ss$ by generalizing the
Sugawara construction. The generator $W_{n}^{\s'}$ with $\s'$-grade $mN$
\cite{WAL} is a differential operator in $x$:
\EQ{
W_{n}^{\s'} \equiv W_{n}^{\s'} \left({\partial\over \partial x},
x\right).
\label{Wn}
}
Following the analogy with the Virasoro constraints, let us consider the
following constraints acting on the highest weight vector
\EQ{
R_n \cdot v_{\s} = \Theta W_{n}^{\s'} \left({\partial \over \partial x},
x + t \right)\Theta^{-1} \cdot v_{\s} =0\,.
\label{Wconst}
}
Now, we can calculate the time evolution of the constraint
\EQ{
{\partial R_n \over \partial t_j} \cdot v_{\s} & =
\left\{ \Theta {\partial \over \partial t_j}
W_{n}^{\s'} \left({\partial \over \partial x}, x + t \right)
\Theta^{-1}\right.  \\ 
&\qquad\quad - \left.\left[ \Pneg(\Theta b_j \Theta^{-1}) ,
\Theta W_{n}^{\s'} \left({\partial \over \partial x},
x + t \right)\Theta^{-1} \right] \right\} \cdot v_{\s}   \\ 
& =  \left[ \Ppos\left( \Theta b_j \Theta^{-1}\right) ,
R_n \right]\cdot v_{\s} =0\,,  
\label{cuenta}
}
for any $n$,
where we have used \eqref{trick}, \eqref{Wconst}, and the fact that $v_{\s}$ is
annihilated by $\gg_{>0[\s]}$, and an eigenvector of
$\gg_{0[\s]}$.
Obviously, the reason for the consistency of \eqref{Wconst} with the
hierarchy is, as in the Virasoro constraints case, the identity
\EQ{
\left[ {\partial \over \partial t_j } - b_j ,
W_{n}^{\s'}\left( {\partial\over \partial x}, x+ t\right)\right] =0\,,
}
and, of course, we could add arbitrary constant elements of $\ss$.

The constraints \eqref{Wconst} induce $\cal W$-constraints on the
tau-functions, in particular
\EQ{
W_{n}^{\s'} \left( {\partial\over \partial x}, x\right)\cdot \tau_{\s} =0\,,
\label{Wcongen}
}
which should be written in terms of the elementary tau-functions.

Before making the obvious conjecture, let us first comment on the
$sl(n)^{(1)}$ case. In this case the relevant $\cal W$-algebra
is the classical ${\cal W}_{n}$, which has generators $W_{k}^{(m)}$,
with $k \in {\Bbb Z}$ and
$m=2,\ldots,n$ ($W_{k}^{(m)}$ can be written as the ordered product of
$m$ elements of $sl(n)^{(1)}$). The resulting
W-constraints are thought to be:
\EQ{
W_{k}^{(m)}\cdot \tau_0 = 0 \qquad k\geq 1-m\,,
}
in the KdV case \cite{WCONST}, and
\EQ{
W_{k}^{(m)}\cdot \tau_i = 0 \qquad k\geq 0\,,
}
for $s_i\neq 0$, in the (p)mKdV case \cite{POOP}.

The natural conjecture is that one should impose ${\cal W}$-algebra constraints
associated to the ${\cal W}$ algebra construct from the twisted scalar field in
\eqref{TSF}. This has generators $W_k^{(m)}$, with $m-1\in I$. The conjectured
constraints are then
\EQ{
W_k^{(m)}\cdot\tau_0=0\qquad k\geq1-m,\quad m\in I+1.
}
Generalized constraints for the (p)mKdV hierarchies follow in an obvious way.
\section{Discussion}
Our final result is quite simple to state. One can construct a generalized
model corresponding to each vertex operator representation of a Kac-Moody
algebra $\gg$, although we only considered the untwisted affinizations of a
simple-laced Lie algebra. In general the partition function of the model is
the product of tau-functions corresponding to the basic representations of
the algebra, each factor subject to a string equation which takes the form
of a Virasoro constraint. The constraint is either $L_{-1}$ if the partition
function is a single factor, and so the hierarchy is of KdV type, and $L_0$
if the partition function contains more than one factor. The relevant
Virasoro generators are those constructed from the Heisenberg subalgebra of
the Vertex construction. The hierarchy together with the string equation
imply an infinite set of constraints which conventional wisdom says is a
${\cal W}$-algebra.

We now make some brief comments about the possible physical interpretation
of the algebraic structure that we have discussed.
Firstly, the Virasoro constraints are
interpreted in the standard way as recursion relations for the correlation
functions of the model. This fixes the so-called ``dilaton equation'' which
specifies the genus expansion of the model. One of the most interesting
questions concerns the existence of new topological models: for instance, a
new topological model was found in \cite{TAM} and discussed more fully in
\cite{PAS} by looking at a complex version of the non-linear Schr\"odinger
hierarchy. At a
``topological point'', in the moduli space, the ${\cal W}$-algebra constraints
determine all the correlation function algebraically. We have found that the
hierarchy based on $A_2^{(1)}$ and the ``intermediate'' Heisenberg
subalgebra (the
hierarchy whose second Hamiltonian structure is the $W_3^{(2)}$-algebra
\cite{GEN2}) leads to a topological model, in the sense of above, although
as in the case of the non-linear Schr\"odinger hierarchy we do not know its
interpretation.

We do not know yet whether the generalized models can be derived from a
matrix model. The only cases which have been analyzed in detail are the
hierarchies associated to $\gg=A_1^{(1)}$ \cite{US}. In this case there
are two
Heisenberg subalgebras: the principle and the homogeneous. The former which
has $\s'=(1,1)$, leads to the original KdV hierarchy, if $\s=(1,0)$, and
the original mKdV hierarchy if $\s=(1,1)$. The latter which has $\s'=(1,0)$
leads to a complex version of the non-linear Schr\"odinger hierarchy. It has
been shown that all the $A_1^{(1)}$ hierarchies can arise from the
one-matrix model by making different double scaling limits. The string
equations that are found in \cite{US} are precisely the ones
constructed in this paper.

There have been other approaches to matrix models which make some claim to
resolve the problem of the non-perturbative ambiguities in the conventional
matric model, whilst preserving an integrable structure \cite{POOP}. This
work falls within the scope of our formalism, one simply takes instead
of a KdV hierarchy the corresponding mKdV hierarchy, and performs the Miura
map, which is trivial at the level of the tau-functions. As we have shown a
mKdV hierarchy admits a different string equation than the associated KdV
hierarchy and it is this difference which gives the theory a unique
non-perturbative definition.
\section*{Acknowledgements}
JLM would like to thank P.G.~Grinevich, S.~Kharchev, A.~Mironov,
A.Y.~Orlov, and A.~Semikhatov for their comments.

\end{document}